# Phonon-assisted optical excitation in the narrow bandgap Mott insulator $Sr_3Ir_2O_7$

H. J. Park,[1,2] C. H. Sohn ,[1, 2] D. W. Jeong,[1,2] G. Cao,[3] K. W. Kim,[4] S. J. Moon,[5] Hosub Jin[1,2], Deok-Yong Cho[1,2,*], and T. W. Noh[1,2]

[1]*Center for Correlated Electron Systems, Institute for Basic Science, Seoul National University, Seoul 151-747, Korea*

[2]*Department of Physics and Astronomy, Seoul National University, Seoul 151-747, Korea*

[3]*Center for Advanced Materials, Department of Physics and Astronomy, University of Kentucky, Lexington, Kentucky 40506, United States*

[4]*Department of Physics, Chungbuk National University, Cheongju 361-763, Korea*

[5]*Department of Physics, Hanyang University, Seoul 133-791, Korea*

## Abstract

We examined the temperature ($T$) evolution of the optical conductivity spectra of $Sr_3Ir_2O_7$ over a wide range of 10–400 K. The system was barely insulating, exhibiting a small indirect bandgap of ≤ 0.1 eV. The low-energy features of the optical $d$-$d$ excitation ($\hbar\omega$ < 0.3 eV) evolved drastically, whereas such evolution was not observed for the O $K$-edge X-ray absorption spectra. This suggests that the $T$ evolution in optical spectra is not caused by a change in the bare (undressed) electronic structure, but instead, presumably originates from an abundance of phonon-assisted indirect excitations. Our results showed that the low-energy excitations were dominated by phonon-absorption processes which involve, in particular, the optical phonons. This implies that phonon-assisted processes significantly facilitate the charge dynamics in barely insulating $Sr_3Ir_2O_7$.

[*]E-mail: zax@snu.ac.kr

## I. Introduction

Iridates have received much attention for their spin-orbit coupling (SOC) effects on various electronic aspects, including metal–insulator transition (MIT) [1-4] and quasiparticle dynamics such as those observed for magnons or polarons [5, 6]. The known mechanisms of the MIT are quite diverse and still controversial; for examples, Mott–Hubbard-type MIT in Ruddlesden–Popper (RP) series iridates ($Sr_2IrO_4$ and $Sr_3Ir_2O_7$) [2] and the topological insulating nature in the honeycomb iridate $(Na/Li)_2IrO_3$ [7]. Polaronic nature and large-energy magnon excitation were also suggested in the case of RP iridates [5, 6]. At the heart of these electronic properties lies the charge excitation across the $J_{\mathrm{eff}}$ quantum states, stabilized by strong SOC in the 5*d* spin-orbitals [8].

This charge excitation is important because its energy scale (a few tenths of eV) offers a glimpse into the charge dynamics and their temperature ($T$)-dependent evolution. In the case of RP-series or honeycomb iridates, the lowest-energy charge excitations occur between two adjacent $J_{\mathrm{eff}}$ = 1/2 states of a lower Hubbard band (LHB) beneath the Fermi level ($E_F$) and an upper Hubbard band (UHB) above the $E_F$, which are split by the on-site Coulomb interactions as well as the SOC. The two bands are more or less parallel to each other with an energy gap of ~0.5 eV [9]. Therefore, these iridate systems can be called “semiconductors with strong electron correlations”.

It is well known that in conventional semiconductors, such as Si or Ge, certain indirect optical excitations constitute the majority of the low-energy optical processes [10, 11]. In an indirect optical process, the electron excitation can be accompanied by absorption or emission of phonons; the momentum of the final electron state can be different from that of the initial state by the momentum value of the phonon involved in the indirect optical process [12]. The indirect bandgap ($E_g$) in Si or Ge is much larger (>1 eV) than the energy scale of the thermal fluctuation. Therefore, the contribution of the indirect optical processes to the optical conductivity can be easily distinguished from any kinds of thermal effects. For example, the Burstein–Moss effects due to thermal occupation (emptiness) of the conduction (valence) bands lead to an increase of energy gap by the order of $k_BT$ ($k_B$: Boltzmann contant) [13, 14]. Since the thermal energy is lower than 0.1 eV in a moderate $T$ range (< 1000 K), this cannot explain the large evolutions of the optical absorption coefficients in the conventional semiconductors.

In contrast, in the case of iridates, $E_g$ is much smaller ($\leq$ 0.1 eV) than Si or Ge, but is comparable with the thermal energies. Under certain phonon-absorbing conditions, the energy cost to excite an electron ($E_g - E_{ph}$; where $E_{ph}$ is phonon energy) can be merely a few hundred Kelvin. In this case, low-energy dynamics can show a significant $T$ dependence. Therefore, in this work, we focused on examining the $T$ dependence of the low-energy optical excitations using optical spectroscopy. To determine the role of phonons, our primary interest, we highlighted the contributions of infrared (IR)-active optical phonon modes on the electronic part in the optical spectra. We also performed O $K$-edge X-ray absorption spectroscopy (XAS) to examine the possible evolution in the bare (undressed) electronic structure.

## II. Experimental Details

High-quality single-crystal $Sr_3Ir_2O_7$ was grown using a self-flux method described elsewhere [15]. The size of the sample was 0.3 mm × 0.3 mm × 0.05 mm. We cleaved the samples to obtain a clean surface for all measurements. We measured the *ab*-plane reflectance spectrum $R(\omega)$ using a near-normal incident beam in the energy region from 0.02–1 eV over a wide $T$ range (10–400 K). We could not obtain the data for energy lower than 0.02 eV because of the small sample size. The complex dielectric functions from 0.74 eV to 5 eV were measured over the same $T$ range using spectroscopic ellipsometry. Then we obtained $\sigma(\omega) = \sigma_1(\omega) + i\sigma_2(\omega)$ from Kramers–Kronig analysis.

Polarization-dependent O $K$-edge XAS experiments at $T$ = 50 K and 400 K were performed at the 2A beamline of the Pohang Light Source (PLS) in total electron yield mode. We obtained the XAS signals from $Sr_3Ir_2O_7$ crystal after cleaving in the vacuum (~1 × $10^{-9}$ torr). In order to check if there are any surface-related effects, we collected the signals with various incidence angles of the beam while maintaining the beam polarization parallel to the sample plane. We did not observe any distinction in spectra in the energy range of interests (Ir 5$d$, <532 eV). Thus, we confirmed that the obtained spectra indeed reflect the bulk properties of the crystal.

## III. Results and Discussion

### A. Optical Spectroscopy: Temperature Evolution of Low-energy Features

Figure 1(a) shows the local density approximation (LDA)+ U band structure of $Sr_3Ir_2O_7$ taken from Ref. [2]. Compared with the case of two-dimensional material $Sr_2IrO_4$, each of the $J_{eff}$ = 1/2 UHB and LHB (shown as blue lines) splits into two due to the electronic interactions between the two perpendicularly adjacent $Ir^{4+}$ ions. The additional splitting of bands lowers the $E_g$ to 85 meV. The transitions $\alpha$ and $\beta$ (shown as vertical arrows) indicate the direct transitions from the $J_{eff}$ = 1/2 LHB to the $J_{eff}$ = 1/2 UHB, and from the $J_{eff}$ = 3/2 bands to the $J_{eff}$ = 1/2 UHB, respectively [2].

Note that the UHB is very flat particularly near the Γ and M points as highlighted by the thick lines in Fig. 1(a). $J_{eff}$ = 1/2 UHB obtained by a dynamical mean field theory (DMFT) calculation are found to be even flatter [16]. The flatness or loss of dispersions indicates significant enhancement of the density of states (DOS). Along with the abundance of optical phonons, which can convey large momentum, the flat UHB near the M point can provide an appreciable probability for phonon-assisted optical excitations from the LHB near the Γ point. It should be noted that in the DMFT calculation result (Ref. 16), the energy of the valence band at Γ point is lower than that at the X point in contrast to the case of the LDA+U calculation (Fig. 1(a)). This finding is consistent with the angle resolved photoemission data [6,8]. Therefore, the low-energy optical transitions can in fact occur from X to M as well as from Γ to M.

Figure 1(b) shows the optical conductivity spectra ($\sigma_1(\omega)$) of $Sr_3Ir_2O_7$ at various $T$. The overall spectra show two optical transitions of $\alpha$ at $\hbar\omega$ ~ 0.35 eV and $\beta$ at $\hbar\omega$ ~ 0.8 eV, consistent with the electronic band structure of $Sr_3Ir_2O_7$ [2, 17]. The sharp spikes that appear below $\hbar\omega$ = 0.1 eV are optical phonon peaks. The peak energy of $\beta$ does not evolve with $T$ suggesting no change in the electronic structure. Meanwhile, the peak energy of $\alpha$ appears to decrease with increasing $T$. This apparent evolution in the peak energy is most probably due to emergence of a low energy feature below $\hbar\omega$ = 0.3 eV. The spectral weights (SWs) of the low-energy features are gradually enhanced with increasing $T$, whereas the main $\alpha$ and $\beta$ features become weakened. As verified in the following, this is a signature of enhancement of phonon-assisted optical excitations at high $T$.

The inset of Fig. 1(b) shows the spectra in a semi-logarithmic plot, which highlights the low-energy region. The value of the $E_g$, roughly estimated by extrapolation at the highest slope of the low-energy feature, was ~0.1 eV for the lowest $T$ (10 K) data. This value is four times smaller than the theoretical direct bandgap of $Sr_3Ir_2O_7$ (~0.4 eV) [2], but is similar to the value of the indirect bandgap (85 meV) shown in Fig. 1(a). This suggests that the optical bandgap observed in Fig. 1(b) is, in fact, the indirect bandgap. The small value of the indirect bandgap suggests the barely insulating nature of $Sr_3Ir_2O_7$. It also reflects that certain optical processes, in which electron momentum is not conserved, should dominate the optical transition in the low-energy region ($\hbar\omega$ < 0.3 eV). The clear $T$ evolution of $\sigma_1(\omega)$ in the low-energy regime suggests that the relevant optical process is significantly facilitated by increasing *T*. Meanwhile, the Drude feature was not observed for all spectra (even at high *T*) despite the increase of the overall spectral weight. This indicates a lack of carriers and a bad-metallic behavior at high *T*.

B. X-ray Absorption Spectroscopy: Bare Electronic Structure

The origin of the *T* evolution in $\sigma_1(\omega)$ can be either a prevalence of indirect optical processes across the $E_g$ or a change in the electronic structure itself. To examine the possibility of the latter (evolution of the bare electronic structure), we performed XAS. Generally, the time scale for incorporation of phonons into the electronic properties is typically on the order of picoseconds. The optical excitation reflects, in good accord, the dynamics of the phonon coupling due to the long time decay process following the photo-excitation. In contrast, XAS has a much shorter time scale (~femtoseconds) for the decay process; the photo-excited electrons decay so fast that they cannot "feel" the phonon coupling during filling up the core holes. Therefore, there is no room for phonon assistance for the electronic transitions in XAS.

Figure 2 shows the O *K*-edge XAS spectra taken at $T$ = 50 K and 400 K. With O *K*-edge XAS, we can probe the unoccupied Ir *d* orbital states that are hybridized with O 2*p* orbitals. We used grazing incidence ($20^o$) measurement geometry with two perpendicular polarizations of $\sigma$ and $\pi$, as depicted in the inset of Fig. 2. The spectra were normalized by the height of the broad Ir *sp* features at higher energy (~535 eV). The huge polarization dependence in the spectra clearly shows the quasi-two-dimensional (2-D) nature of $Sr_3Ir_2O_7$, which stands in the middle of the RP series from $Sr_2IrO_4$ to $SrIrO_3$ [2]. Features can be assigned according to the order of the Ir *d* and O [apical ($O_A$) or planar ($O_P$)] 2*p* hybridized states [18].

The low-energy feature ($\hbar\omega$ ~ 527 eV) in the ($E//\pi$) data ($\pi$: $\sin^2 70^o$ ($E//c$) + $\cos^2 70^o$ ($E//ab$)) is the $J_{eff}$ = 1/2 UHB, which corresponds to the $\alpha$ peak in $\sigma_1(\omega)$ (Fig. 1(b)). The most important finding in this figure is that the intensity of the first peak barely evolves with *T*. This shows that there was almost no change in the (bare) electronic structure. This is in a sharp contrast to the case of $\sigma_1(\omega)$. The overall XAS features, including *yz*/*zx*-$O_P$, ($3z^2$-$r^2$)-$O_{A/P}$, and ($x^2$-$y^2$)-$O_P$, became slightly broadened at higher *T*, possibly due to thermal disorders (thermal vibrational motions of ions).

We noticed that all of the features for both polarizations were shifted slightly, but concurrently, by the same amount (~ −0.05 eV) at high *T* ($T$ = 400 K). The energy shift is presumably caused by a difference in the chemistry of the O ions (chemical shift). In general, valence of ions can change with temperature due to their thermal motions. The thermal motion of the $Ir^{4+}$ and $O^{2-}$ ions can lead to a decrease of the Ir *d*-O *p* hybridization because the lateral displacement of the $O^{2-}$ ions with respect to the $Ir^{4+}$ ions will effectively increase the Ir-O bond length (see Fig. 3(c) for the description of the thermal vibrations). According to Harrison's $a^{-7}$ rule ($a$: interatomic distance) [19], the strength of the Ir *d*-O *p* hybridization should decrease as the effective bond length increases. Weakness in hybridization not only decreases the O *K*-edge XAS intensities slightly but also lowers the energies of the features (by ~-0.05 eV), as was observed in the O *K*-edge XAS spectra in Fig. 2. Because this chemical shift is merely a final state effect in the XAS, it cannot explain the drastic $T$ evolution in the low-energy part of $\sigma_1(\omega)$ (Fig. 1(b)). Therefore, we can conclude that the $T$ evolution in $\sigma_1(\omega)$ cannot originate from the evolution in bare electronic structure.

C. Phonons: Softening of a Bending Mode

Figures 3(a) and 3(b) show three major IR-active phonon modes, which appear below 0.1 eV in $\sigma_1(\omega)$ (Fig. 1(b)). Two phonon modes were located at ~264 $cm^{-1}$ and ~374 $cm^{-1}$ (Fig. 3(a)) and corresponded to the bending modes within the $IrO_6$ octahedron (O-Ir-O bond-angle changes). The highest-energy phonon mode was located at ~639 $cm^{-1}$ (Fig. 3(b)) and corresponded to stretching of the octahedron (Ir-O elongation/contraction). The peak heights of the features were weakened as $T$ increased due to thermal disorder while their SWs were roughly preserved. These three optical phonon modes can contribute to the optical excitation of low $\hbar\omega$ below the indirect bandgap (85 meV) by donating their energy and momentum for electronic excitation. Their minimal dispersions [20] as well as high energies can facilitate the indirect optical excitations at the low $\hbar\omega$ very efficiently.

Figure 3(c) shows the $T$ evolution of the wavenumbers of the three phonon modes. The values were normalized by the respective values at the lowest $T$. The 374 $cm^{-1}$ phonon mode softened significantly (its energy decreased by ~6%) at high $T$, whereas the 264 $cm^{-1}$ and 639 $cm^{-1}$ phonon modes did not. This suggests that the $O^{2-}$ ions can have large lateral displacements with respect to the $Ir^{4+}$ ions at high $T$. The inset figure in Fig. 3(c) depicts the thermal vibration in the Ir-O bonds, corresponding to the bending mode. The Ir-O bond length can increase effectively (exemplarily indicated by red arrow) without any crystallographic evolution. Therefore, the decrease of Ir *d*-O *p* hybridization strength observed in the O *K*-edge XAS spectra (Fig. 2) can be explained by softening of 374 $cm^{-1}$ bending mode.

Although the phonon mode softening with increasing $T$ is consistent with the XAS data (Fig. 2), it cannot be correlated with the drastic $T$ evolution of the low-energy optical excitations (Fig. 1(b)). The decrement of the optical phonon energy was at most 3 meV, so that we could neglect the energy difference from that of the low $T$ (~46 meV). Meanwhile, the presence of the three optical phonon modes can in itself facilitate the low energy optical excitation because the high energy phonons can lower the energy barrier ($\hbar\omega = E_g - E_{ph}$) efficiently. Therefore, the abundance of phonon-assisted indirect optical excitations can be significant in determining the low energy features in $\sigma_1(\omega)$. The $T$ evolution of the phonon-assisted optical excitations can be simulated in terms of the phonon abundances. The results of the simulation are demonstrated in the next section.

D. Simulation of the Indirect Optical Transition

In the phonon-assisted transitions, the photo-excited electrons can either absorb or emit discrete phonon energies. Figure 4(a) schematically depicts the direct and indirect optical transitions between two $J_{\text{eff}} = 1/2$ bands. How the phonons are involved in the optical transition can be described in the matrix element for the indirect processes. If the variation in the matrix element can be neglected, we may assume that the prevalence of the indirect processes is roughly proportional to the number of phonons multiplied by the given DOSs of the initial and final electronic states at a particular $\hbar\omega$.

The number of phonons can be described in terms of Bose–Einstein statistics: $1/(\exp(E_{ph}/k_BT) - 1)$, namely, a lower phonon energy ($E_{ph}$) means a larger number of phonons, and therefore, a larger probability of indirect optical processes. Thus, in general, acoustic phonons are incorporated with indirect transition more dominantly than optical phonons due to their higher population at low energy. However, the contribution of acoustic phonons particularly in the low-$\hbar\omega$ features would be less significant because their $E_{ph}$'s are overall smaller than those of the optical phonons. The value of the threshold energy for the onset of optical absorption is given as $\hbar\omega = E_g - E_{ph}$ for the phonon absorption process and $\hbar\omega = E_g + E_{ph}$ for the phonon emission process. This suggests that the low-$\hbar\omega$ optical excitations are dominated by phonon-absorbing processes.

Based on the three major optical phonon modes, the phonon-absorbing/emitting optical processes between the $J_{\text{eff}} = 1/2$ LHB and UHB were simulated for various $T$. In the calculation, we used the LDA+U DOS of $Sr_3Ir_2O_7$ from Ref. [2] and neglected the evolution of the phonon energy itself. Here, all of the matrix elements were assumed to be constant. The results are shown in Fig. 4(b). When $\hbar\omega$ is much smaller than the direct bandgap, the optical conductivity can be written approximately [12] as

$$\sigma_1(\omega,T) = \sum_{\text{phonons}}\left[\frac{1}{\omega}\int dE\, DOS_i\,(E)\cdot DOS_f\left(E+\hbar\omega+E_{ph}\right)\frac{1}{\exp\left(\frac{E_{ph}}{k_BT}\right)-1}+\frac{1}{\omega}\int dE\, DOS_i\,(E)\cdot DOS_f\left(E+\hbar\omega-E_{ph}\right)\frac{1}{1-\exp\left(\frac{-E_{ph}}{k_BT}\right)}\right]$$

, where $DOS_i$ ($DOS_f$) refer to the $J_{eff}$ = 1/2 LHB (UHB) DOS. The former part in the parenthesis is from the phonon-absorption process, and the latter part is from the phonon-emission process. It should be noted that this equation holds only at low $\hbar\omega$, namely, in the region where the phonon-absorption process dominates (≤ 0.1eV).

As $T$ increases, the overall indirect transition is enhanced due to the drastic increase in the number of optical phonons (i.e., due to Bose–Einstein statistics). Three abrupt edge-jumps were present, according to the onsets of phonon-absorbing optical processes. Because the highest energy of the phonon modes was ~80 meV (for 639 $cm^{-1}$) and $E_g$ = 85 meV, the onset energy of the indirect transition was ~5 meV; this value is quite small and comparable to the thermal energy at room temperature. Thus, we may expect uncertainty in the spectral function including a substantial broadening of features due to thermal vibrations of the ions. It is well known that the vibrational motions of ions in solid are quantized as to yield discrete energy distribution above the electronic ground state. This will induce some variation in the energy of the optical transition as to broaden the overall features in the spectra. Especially for low $\hbar\omega$ region, the bandgap can appear to be closed at high $T$. The low energy part in the experimental $\sigma_1(\omega)$ in Fig. 1(b) indeed showed a $T$ evolution closing the gap. This shows that the experimental $\sigma_1(\omega)$ is consistent with the simulation result, although the step-like features are largely smeared out presumably due to the thermal disorders.

Therefore, it can be told that the phonon-assisted optical excitation dominates the low energy electronic excitations in barely insulating $Sr_3Ir_2O_7$. Its contribution was maximized under the abundance of high-energy optical phonons. The flatness of the UHB (particularly near the M point) also contributed to the enhancement of the indirect processes since it increases the $DOS_f$ significantly. The phonon-assisted transition for $\beta$, was supposed to be buried under the energy region of the direct transition $\alpha$. Its contribution might be relatively weak because all of the four $J_{eff}$ = 3/2 subbands had quite different dispersions from the $J_{eff}$ = 1/2 UHB, in contrast to the case of nearly parallel $J_{eff}$ = 1/2 LHB.

## IV. Outlook and Conclusion

This work demonstrates the importance of indirect optical processes for the charge dynamics of narrow indirect-bandgap 5*d* insulators. This is the reminiscent of the physics of conventional semiconductors, such as Si or Ge, although it becomes more complicated due to much smaller $E_g$'s. The dominant role of phonon-assisted charge excitations suggests future work to understand the underlying physics in 5*d* transition-metal oxides (TMOs). At this moment, we cannot judge with this data if the *T* dependence in the optical conductivity can be incorporated with the Mott- or Slater-type MIT scheme. Charge excitation is dominated by the phonon-assisted transition across the Mott gap between the two $J_{\text{eff}}$ = 1/2 bands, while the AFM order can also alter the electronic structure of the $J_{\text{eff}}$ = 1/2 bands and consequently influence the indirect transition. Also, the crude approximation of the constant matrix element used in this work, should be revised in the future because the physics of the relativistic $J_{\text{eff}}$ states dictates that the angular momentum and the transition matrix can vary point-by-point within a band. Thus, the matrix element can vary depending on which phonon mode is involved in the optical process, necessitating the theoretical consideration of symmetry arguments for spin-orbital physics in 5*d* TMOs.

In conclusion, the drastic *T*-dependence in the low-energy *d*-*d* optical excitation in the barely Mott-insulating $Sr_3Ir_2O_7$ suggests the dominance of phonon-assisted optical processes, without a change in the bare electronic structure. It was conclusively shown that optical phonons determine the low-energy charge dynamics with the aid of a small bandgap (≤ 0.1 eV) and flat $J_{\text{eff}}$ = 1/2 upper band.

Acknowledgment

This work was supported by the Research Center Program of the Institute for Basic Science in Korea. Experiments at the PLS were supported in part by MSIP and POSTECH.

Figure Captions

Figure 1: (color online) (a) Electronic structure of $Sr_3Ir_2O_7$ (LDA+U) taken from Ref. [1]. The direct transitions $\alpha$ and $\beta$ are indicated exemplarily. The indirect bandgap ($E_g$) is estimated to be 85 meV. (b) The experimental $\sigma_1(\omega)$ of $Sr_3Ir_2O_7$ at various $T$. The low energy features ($\hbar\omega$ < 0.3 eV) are enhanced with increasing $T$. Inset: a semi-logarithmic plot to highlight the evolution of the low energy features.

Figure 2: (color online) O *K*-edge XAS spectra taken with two perpendicular polarizations of $\sigma$ and $\pi$ at $T$ = 50 K and 400 K. The measurement geometry is depicted in the inset.

Figure 3: (color online) Three major optical phonon modes: (a) bending modes in the $IrO_6$ octahedron at 264 $cm^{-1}$ and 374 $cm^{-1}$, and (b) stretching mode for Ir-O bond. (c) The $T$ evolution of the phonon energies normalized by the lowest $T$ ($T$ = 10K) data. Softening of the 374 $cm^{-1}$ phonons suggests the increase of the average Ir-O bond length and the decrease of orbital hybridization due to thermal vibrational motions of the ions.

Figure 4: (color online) (a) Schematic of the optical excitations including direct photo-absorption, phonon-absorbing or phonon-emitting indirect photo-absorptions. (b) The results of the simulation for the phonon-assisted optical absorptions based on actual $J_{eff}$=1/2 LDA+U band structure of $Sr_3Ir_2O_7$.

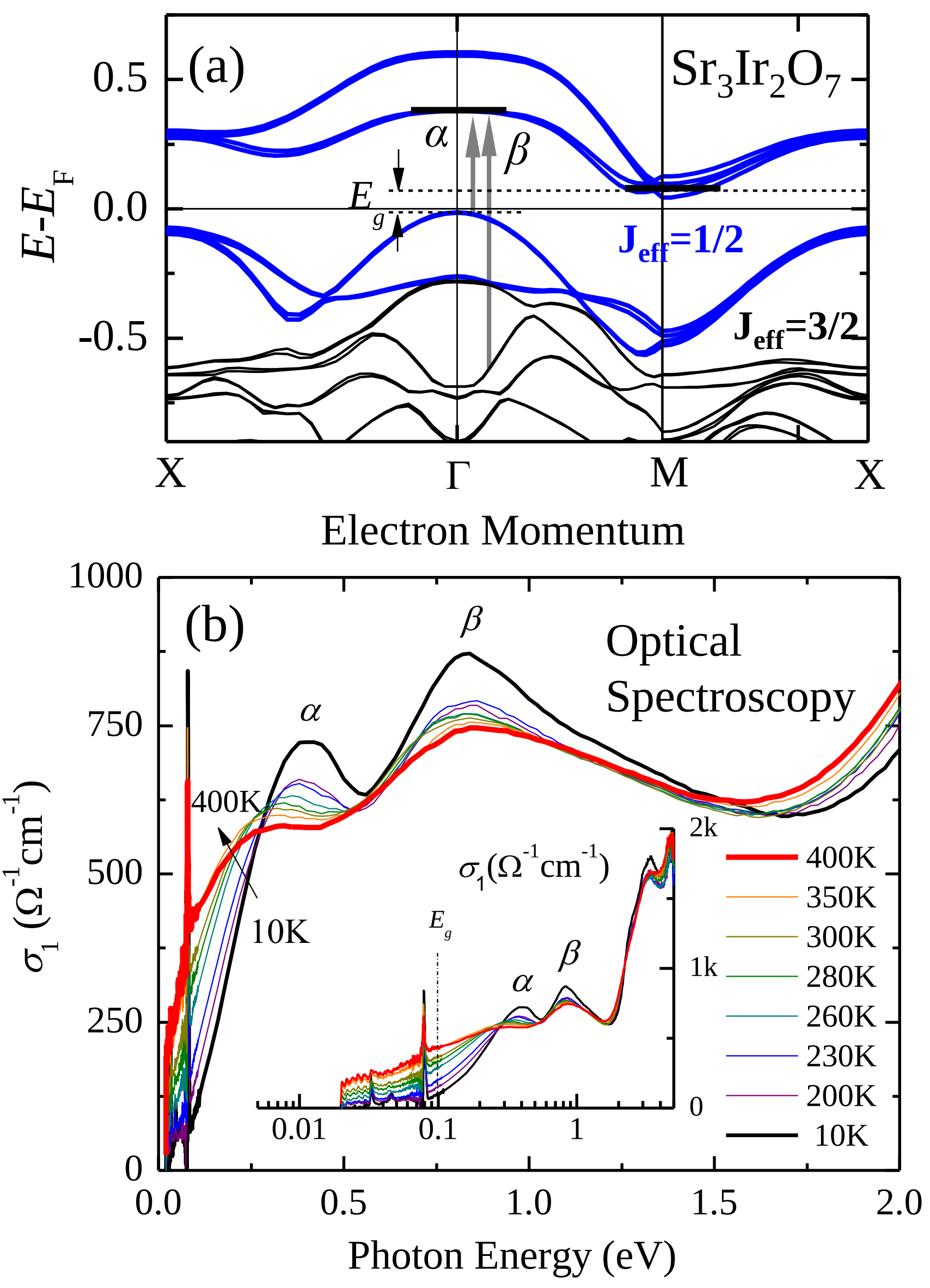

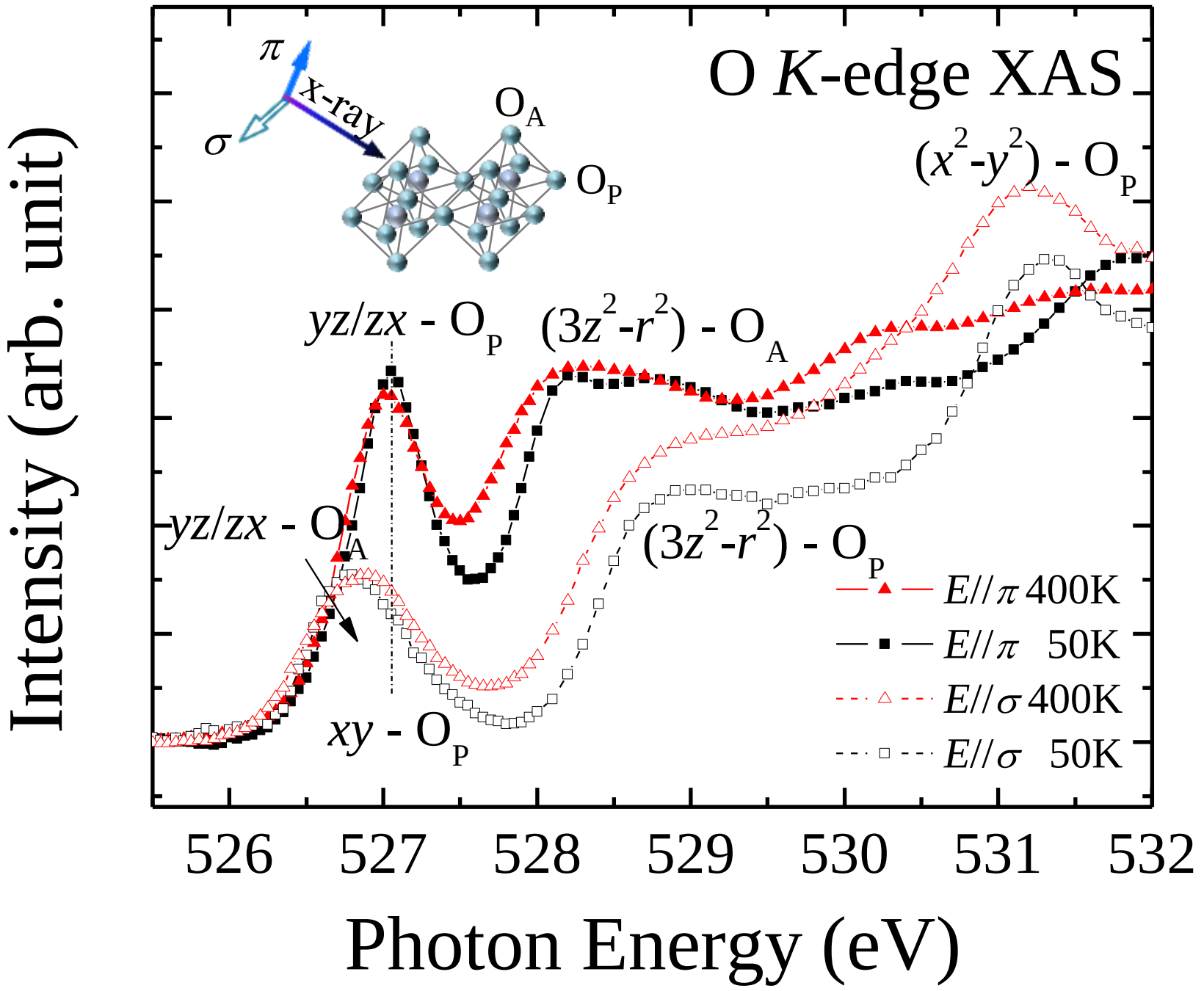

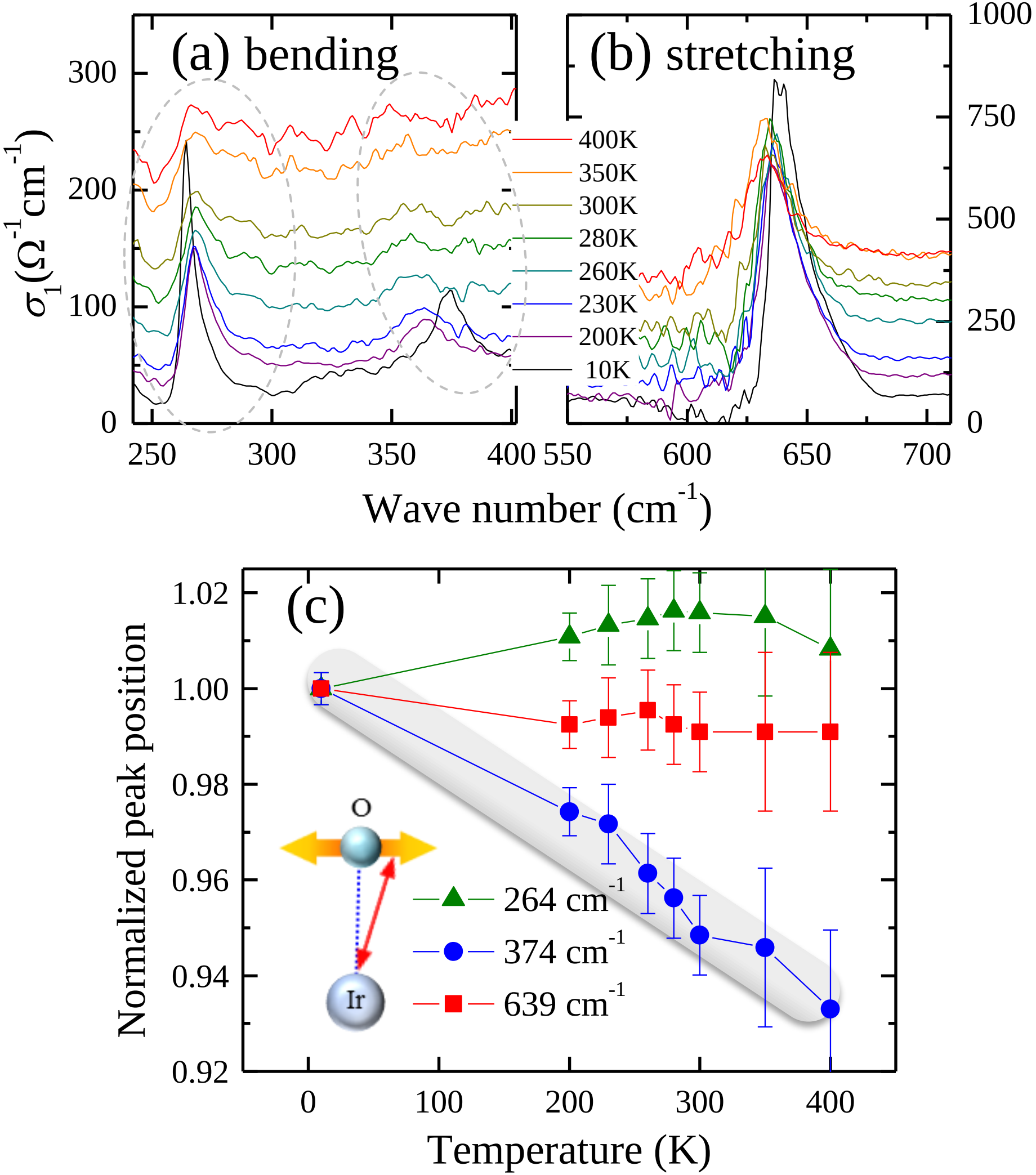

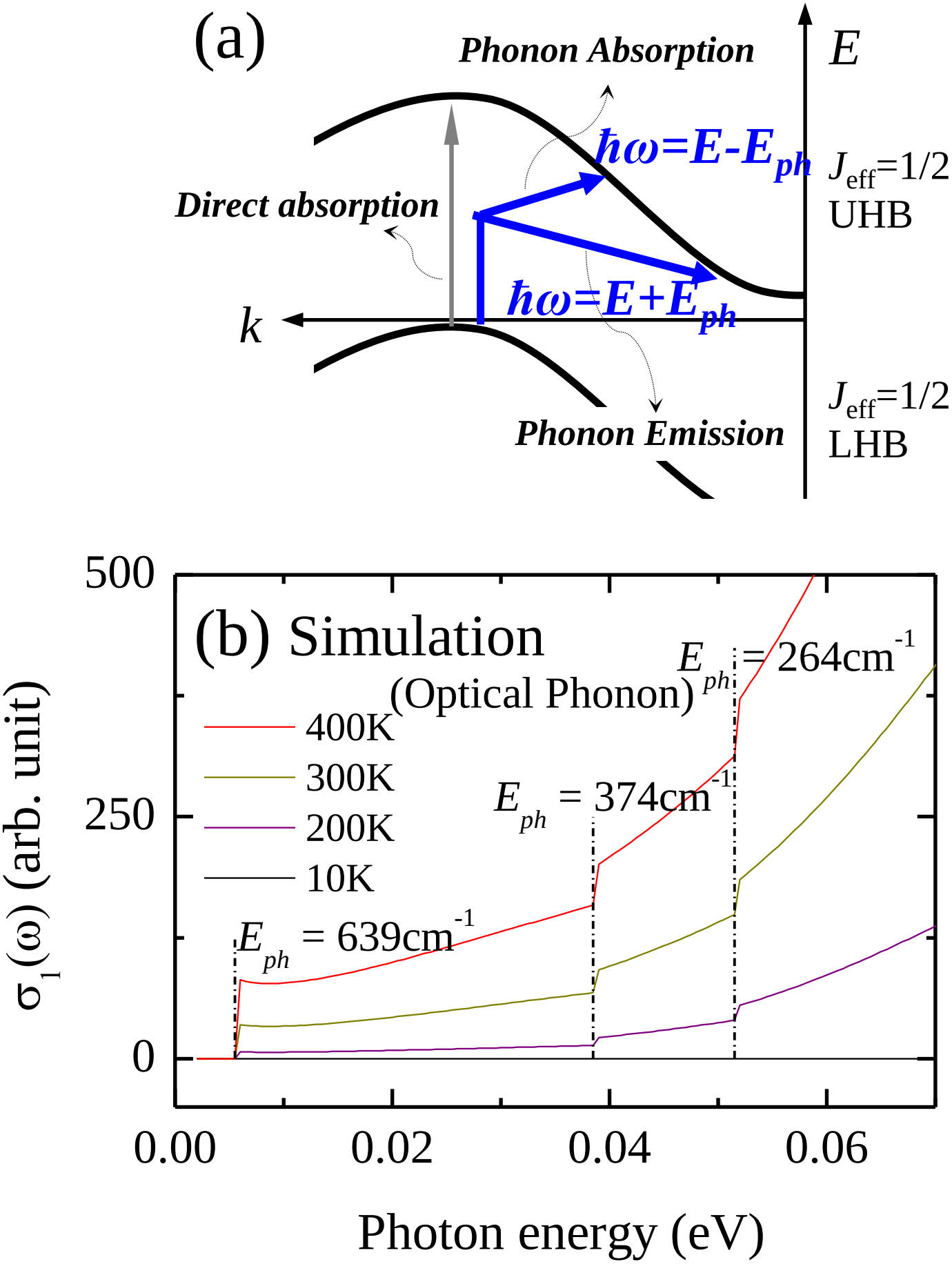